# Negative longitudinal magnetoresistance from anomalous N=0 Landau level in topological materials

B.A. Assaf[1], T. Phuphachong[2], E. Kampert[3], V.V. Volobuev[4,5], P.S. Mandal[6], J. Sánchez-Barriga[6], O. Rader[6], G. Bauer[4], G. Springholz[4], L.A. de Vaulchier[2], Y. Guldner[2]

[1] *Département de Physique, Ecole Normale Supérieure, PSL Research University, 24 rue Lhomond, 75005 Paris, France*

[2] *Laboratoire Pierre Aigrain, Ecole Normale Supérieure, PSL Research University, Université Pierre et Marie Curie, Sorbonne Universités, Université Denis Diderot, Sorbonne Cité, 24 rue Lhomond, 75005 Paris, France*

[3] *Dresden High Magnetic Field Laboratory (HLD-EMFL), Helmholtz-Zentrum Dresden-Rossendorf, 01328 Dresden, Germany*

[4] *Institut für Halbleiter und Festkörperphysik, Johannes Kepler Universität, Altenberger Straße 69, 4040 Linz, Austria*

[5] *National Technical University "Kharkiv Polytechnic Institute", Frunze Str. 21, 61002 Kharkiv, Ukraine*

[6] *Helmholtz-Zentrum Berlin für Materialien und Energie, Albert-Einstein Str. 15, 12489 Berlin, Germany*

**Abstract:** Negative longitudinal magnetoresistance (NLMR) is shown to occur in topological materials in the extreme quantum limit, when a magnetic field is applied parallel to the excitation current. We perform pulsed and DC field measurements on $Pb_{1-x}Sn_xSe$ epilayers where the topological state can be chemically tuned. The NLMR is observed in the topological state, but is suppressed and becomes positive when the system becomes trivial. In a topological material, the lowest N=0 conduction Landau level disperses down in energy as a function of increasing magnetic field, while the N=0 valence Landau level disperses upwards. This anomalous behavior is shown to be responsible for the observed NLMR. Our work provides an explanation of the outstanding question of NLMR in topological insulators and establishes this effect as a possible hallmark of bulk conduction in topological matter.

The emergence of topological insulators (TI) as novel quantum materials [1] [2] [3] has played a key role in the discovery of novel physical phenomena, [4] [5] [6] [7] [8] [9] such as the quantum spin Hall effect [4] [10] [11] and the quantum anomalous Hall effect [5] [12] [13]. This stems from the helical Dirac nature of surface-states in 3D TIs or, that of edge-states in 2D TIs. In fact, a huge amount of literature (for reviews [14] [15] [16] [17]) took interest in this question and investigated electronic transport of 2D Dirac electrons in 3D-TIs. The majority of these studies were, however, impeded by the significant and dominant bulk transport that occurs in TIs. On the other hand, little attention has been given to signatures of non-trivial band topology in 3D electron transport in a TI.

Naively speaking, one can think of the bulk energy bands of a TI as being identical to those of conventional semiconductors and, thus, unlikely to generate non-conventional physical phenomena. However, one should not forget that the basis of a topological insulator lies in the inverted orbital character of these bulk energy bands. [10], [18] Most interesting is the unusual behavior of the Landau levels of TIs that one can analytically extract from a general Bernevig-Hughes-Zhang Hamiltonian (appendix of [18]). In fact, it has been both theoretically [18] [19] [20] and experimentally [21] [22] [23] shown, that the energy of the lowest (N=0) conduction (valence) Landau level in topological insulators decreases (increases) as a function of increasing magnetic field, opposite to what usually happens in a topologically trivial system (Fig.1(a,b)). This behavior is anomalous and leads to a field-induced closure of the energy gap in a TI [21] (Fig.1(b)), whereas in a trivial material, the energy gap usually opens with increasing magnetic field (Fig.1(a)). This anomaly is a hallmark of the inverted band structure of topological materials. Its implications on magnetotransport have not yet been considered.

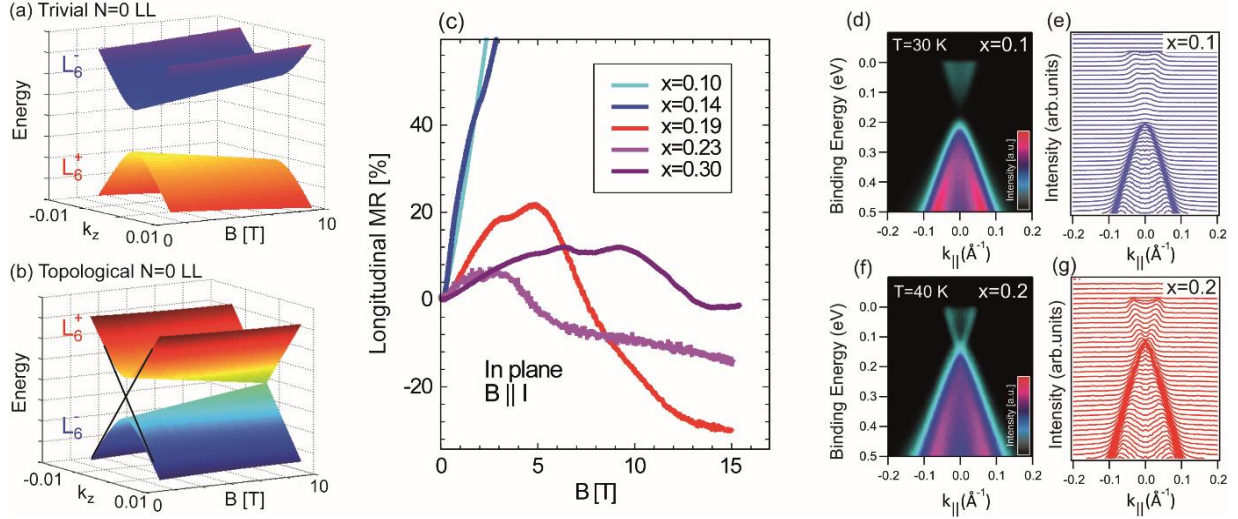

FIG 1. (Color Online) Sketch of the behavior of the N=0 bulk LL as a function of magnetic field for a trivial (a) and a topological (b) system. The k-dispersion of the energy level in the direction of the applied field is shown ($k_z$). Topological surface states are shown in black in (b) at B=0. The $L_6^{\pm}$ bands denote the band extrema of opposite parity occurring at the L-point in $Pb_{1-x}Sn_xSe$. (c) In-plane-MR measured with B || I at 10K in two trivial $Pb_{1-x}Sn_xSe$ epilayers (x=0.10 and x=0.14) and three topological ones (x=0.19, x=0.23 and x=0.3) up to B=15T. ARPES dispersions and momentum distribution curves for x=0.1 (d,e) and x=0.2 (f,g) measured with 18eV photons at 30K and 40K, respectively.

In the present work, we study the MR in topological insulators in the extreme quantum limit – the regime where only the lowest Landau level (LL) is occupied. We measure magnetotransport in pulsed magnetic field up to 61T in high mobility $Pb_{1-x}Sn_xSe$ epitaxial layers. We show that, when all Lorentz contributions to the MR are suppressed by applying the magnetic field in-plane and parallel to the excitation current, a negative longitudinal MR (NLMR) emerges near the onset of the quantum limit. This NLMR is only observed in the topological regime of $Pb_{1-x}Sn_xSe$ (x>0.16) and is absent in trivial samples (x<0.16). We theoretically argue that this NLMR is a result of the anomalous behavior of the N=0 LL that leads to a field induced closure of the energy gap as a function of the applied magnetic field, thus enhancing the carriers' Fermi velocity and reducing electrical resistivity. Our findings establish that NLMR is a hallmark of the topological insulating state, and may reconcile controversial interpretations of axial anomaly-induced NLMR in such materials.

Magnetotransport measurements are performed on [111]-oriented $Pb_{1-x}Sn_xSe$ epilayers grown on (111) $BaF_2$ substrates with different x. Growth by molecular beam epitaxy and characterization are described in our previous works [24] [25] [26]. A 15T/4.2K superconducting cryostat setup is used for in-house measurements. Further measurements are performed at 10K up to 61T using a 200ms pulsed-field coil at the Dresden High Magnetic Fields Lab. Angle-resolved-photoemission (ARPES) experiments are performed with linearly-polarized undulator radiation at the UE112-PGM1 beamline of the synchrotron BESSY-II in Berlin.

Figure 1(c) shows the longitudinal MR measured at 10K, up to 15T for five $Pb_{1-x}Sn_xSe$ samples, with the magnetic field applied in-plane parallel to the current (I//B//[1-10]) (Fig. S2). For trivial samples having x<0.16, [24] the MR rises fast. In non-trivial samples having x>0.16, [24] [27] [28] although initially positive, the MR turns negative, and remains so over a wide range. The sign of the MR hence depends on the topological character of the sample.

ARPES measurements [Figs. 1(d)-1(g)] for x=0.10 and x=0.20 below 50K clearly indicate the changing topological character across x=0.16. A gapped state is observed in the ARPES dispersion and momentum distribution curves for x=0.10 (Fig.1(d,e)) whereas for x=0.20 a gapless topological Dirac surface state is clearly resolved (Fig.1(f,g)), in agreement with previous ARPES studies. [27] [28]. This ties the occurrence of the NLMR to the topologically non-trivial regime in $Pb_{1-x}Sn_xSe$.

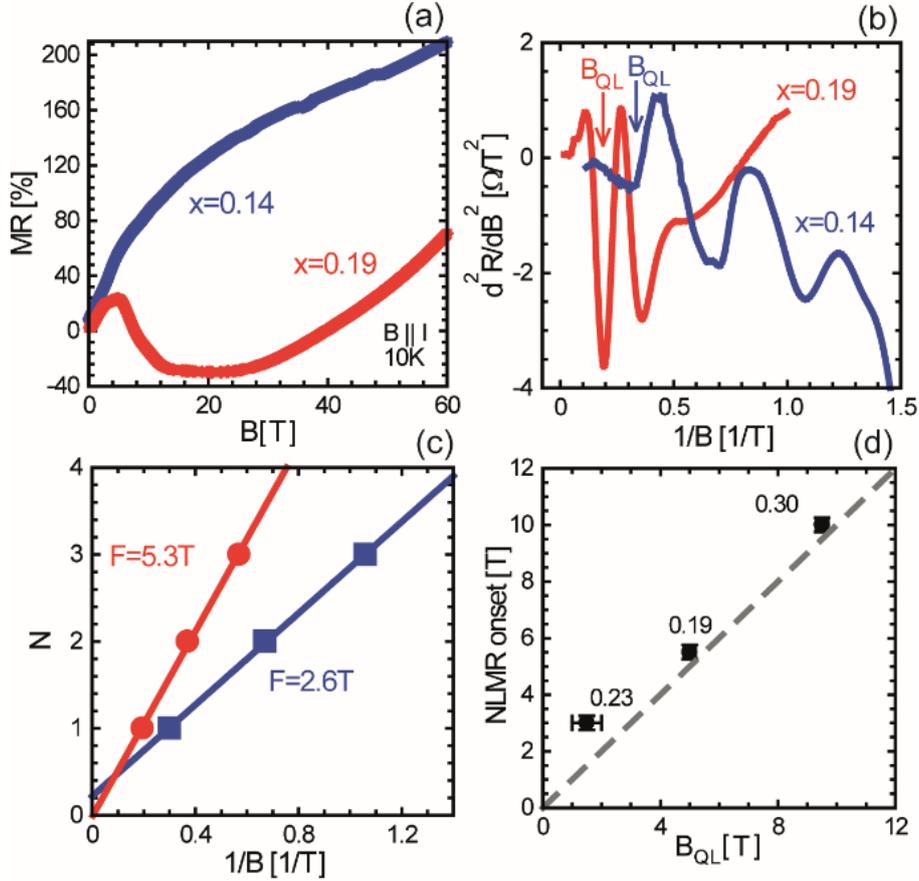

FIG 2. (Color Online). (a) In-plane MR measured up to 60T using pulsed magnetic field for $Pb_{1-x}Sn_xSe$ with x=0.14 (blue) and x=0.19 (red). (b) Low-field Shubnikov-de-Haas oscillations and (c) Landau index versus 1/B shown for both samples. Arrows mark the field at which the quantum limit is reached ($B_{QL}$). (d) NLMR onset extracted from Fig. 1(c) versus $B_{QL}$ for the three topological samples considered in this work. The dashed grey line is obtained for an onset exactly equal to $B_{QL}$. The Sn concentration 'x' corresponding to each sample is shown above the data points.

In order to confirm the robustness of the MR trend on either side of the topological phase transition, transport measurements for fields up to 61T are performed on two selected samples with compositions close to the transition. Results are shown in Fig. 2(a). Comparing the sample x=0.14 to x=0.19 confirms that the MR in the trivial regime is robustly positive up to 60T, whereas in the topological regime, the MR is initially positive, then turns negative and reaches a plateau-like behavior at intermediate fields, then increases again at very high fields.

We correlate the appearance of the NLMR to the crossing of the N=1 LL with the Fermi energy ($E_f$), by looking at 3D Shubnikov-de-Haas (SdH) oscillations measured in the same geometry as the MR (I//B//[1-10]). Fig.2(b) shows SdH oscillations in the second derivative of the resistance for x=0.14 and x=0.19 at 10K. The last oscillation minimum is observed at $B_{QL}\approx 5T$ ($0.2T^{-1}$) for x=0.19 and 2.8T ($0.35T^{-1}$) for x=0.14; this is the onset of the extreme quantum limit (arrows in Fig.2(b)). The SdH frequency extracted from the plot of the Landau index N versus 1/B (Fig.2(c)) comes out close to 5T for x=0.19 and 2.6T for x=0.14. For x=0.19, this yields a 3D carrier density of about $6\times 10^{16} cm^{-3}$ per valley or a total of $2.4\times 10^{17} cm^{-3}$ for the four valleys of $Pb_{1-x}Sn_xSe$. This also agrees with $n_{Hall}\approx 3\times 10^{17} cm^{-3}$. [20] For x=0.14, we find $2\times 10^{16} cm^{-3}$ per valley. The Hall data yields $p_{Hall}=1\times 10^{17} cm^{-3}$ for four valleys in agreement with SdH data. We also note that the SdH results nicely agree with our previous magnetooptical measurements on the same samples. [24] Even though the two samples studied here in detail have different carrier type, the other samples examined in Fig.1(c) rule out any link between this and the NLMR. [20]

In x=0.19, $B_{QL}$ is close to the onset of the NLMR seen in Fig.2(a). In x=0.14, even though $B_{QL}$ is small, no NLMR is observed up to 60T. We consolidate the relation between the NLMR and the entrance into the quantum limit in the topological state by further investigating two additional samples (x=0.23 and x=0.3). Detailed SdH and magnetooptical IR spectroscopy data shown in the supplement allow us to extract $B_{QL}$ for both. [20] The onset of the NLMR extracted from Fig.1(c) is plotted versus $B_{QL}$ for x=0.19, x=0.23 and x=0.3 in Fig.2(d). A clear correlation of the onset of NLMR with $B_{QL}$ is observed, as indicated by the dashed line, confirming that the NLMR occurs in the quantum limit.

We next elucidate the origin of the NLMR occurring in topological materials in the quantum limit by investigating transport in this regime. We have shown that the LL in IV-IV TCIs can be well described by a massive Dirac spectrum that includes spin-splitting [24] [29] [30] [31], resulting from a 6-band **k.p** Hamiltonian. [30] If the contributions from the far-bands are neglected a 2-band **k.p** Hamiltonian results, the solution of which is an ideal massive Dirac model [32].

We use the 6-band Mitchell-Wallis Hamiltonian to describe the field dependence of the N=0 level and its wavevector dispersion [20] [30] [33] [34] [35] [36]:

$$E_0 = \sqrt{\left(|\Delta| \pm \left(\frac{\hbar\tilde{\omega}}{2} - \frac{\tilde{g}\mu_B B}{2}\right)\right)^2 + (\hbar v_z k_z)^2} \qquad (1)$$

The ± sign refers to the trivial and topological regime respectively. $\Delta$ is the half-band-gap, $k_z$ is wavevector and $v_z$ is the Dirac velocity in the z-direction (z//B). $\tilde{\omega} = eB/\tilde{m}$. $\tilde{m}$ and $\tilde{g}$ are the mass and effective g-factor terms resulting from interactions between the band edges, and far-bands located about 1eV above and below the energy-gap in IV-VI semiconductors. [30][31] We highlight, that the Mitchell-Wallis Hamiltonian is similar to the Bernevig-Hughes-Zhang (BHZ) Hamiltonian [18] [10] [37] that generally describes topological systems. Our treatment (Eq. (1)) can thus be generalized to any topological system exhibiting the N=0 behavior shown in Fig.1(b). The $\tilde{m}$ contribution also appears on the diagonal of the BHZ Hamiltonian as $-M_1 k^2 = \frac{\hbar^2 (k_x + k_y)^2}{2\tilde{m}}$. [10] [37]

For PbSe, $\tilde{g}\mu_B B \approx -\hbar\tilde{\omega}$ ( $\tilde{g} \approx \frac{2m_0}{\tilde{m}}$ ). [38] Far-band terms vary little (<10%) with Sn content up to about x= 0.3, according to laser emission measurements in magnetic fields. [21] Using this result, we simplify Eq.(1) to:

$$E_0 = \sqrt{(|\Delta| \pm \hbar\tilde{\omega})^2 + (\hbar v_z k_z)^2} \qquad (2)$$

In Eq. 2, the $|\Delta| \pm \hbar\tilde{\omega}$ term describes whether the energy-gap closes (–) or opens (+) as a function of increasing magnetic field. At very high fields such that $|\Delta| < \hbar\tilde{\omega}$, for both the topological and trivial regime, the energy gap opens with increasing field and the N=0 LL varies as given in Eq. 2 for the (+) case. [20] The field-induced gap closure in the topological regime, is most likely accompanied by a topological phase transition. This effect has been treated theoretically for the QSH state in 2D [39] but not yet for 3D-TIs and TCIs.

The LL energies are plotted versus magnetic field in Fig.3(a,b) for x=0.19 and x=0.14, respectively, (and in the supplement for x=0.23 and x=0.3). The parameters are given in the caption and in table I. For

N=0, Eq.(2) is used. Notice that the N=0 level is non-spin degenerate in both the topological and trivial regimes, carriers are thus fully spin-polarized in the quantum limit in $Pb_{1-x}Sn_xSe$. The magnetic-field dependence of $E_f$ is plotted in Fig.3(a,b) using:

$$n_{SdH} = \frac{eB}{4\pi^2\hbar} \sum_N \int f(E_N, k_z) dk_z \quad (3)$$

$n_{SdH}$ is the valley carrier density, $f(E_N, k_z)$ is the Fermi-Dirac function, $E_N$ is the LL energy. From Eq.(3), we also get $k_z(B)$ in the quantum limit:

$$k_z(B) = \frac{2\pi^2 \hbar n_{SdH}}{eB}, \quad (4)$$

The magnetoconductivity for a 3D electron gas in the quantum limit, in the presence of point-like impurities, has recently been treated by Goswami et al. [40] Although ref. [40] also treated the problem of scattering by long-range ionized impurities, we neglect their impact in $Pb_{1-x}Sn_xSe$ because of its very large dielectric constant (>280). [41] [42] In IV-VI systems, the scattering rate from ionized impurities is thus expected to be at least two-orders of magnitude smaller than that of narrow-gap III-V or II-VI materials. [40] [43] It is also well known that in $Pb_{1-x}Sn_xSe$, doping is essentially caused by atomic vacancies that can be treated as point-like defects. From ref. [40] we get:

$$\sigma(B) = \frac{e^2 \hbar}{2\pi n_i U_0} v_f^2(B) \quad (5)$$

$n_i$ is the impurity density, $U_0$ is the impurity potential and $v_f(B)$ is the Fermi velocity as a function of magnetic field. Using Eq.(2) and (4), we obtain [40]:

$$v_f(B) = \frac{\alpha v_z^2}{\sqrt{\left(|\Delta| \pm \frac{\hbar eB}{\tilde{m}}\right)^2 e^2 B^2 + \alpha^2 v_z^2}} \quad (6)$$

$\alpha = 2\pi^2 \hbar^2 n_{SdH}$. By plugging Eq.(6) into Eq.(5), we get the MR and its derivative in the quantum limit:

$$MR = \frac{\rho(B)}{\rho(0)} - 1 = \frac{\sigma(0)}{\sigma(B)} - 1 = \left(|\Delta| \pm \frac{\hbar eB}{\tilde{m}}\right)^2 \frac{e^2 B^2}{\alpha^2 v_z^2} \quad (7)$$

$$\frac{dMR}{dB} = 2 \frac{e^2 B}{\alpha^2 v_z^2} \left(|\Delta| \pm \frac{\hbar eB}{\tilde{m}}\right) \left(|\Delta| \pm 2\frac{\hbar eB}{\tilde{m}}\right) \quad (8)$$

For $\Delta \tilde{m}/2\hbar e < B < \Delta \tilde{m}/\hbar e$ the derivative becomes negative in the topological regime and resistance decreases as a function of magnetic field, yielding a NLMR.

Qualitatively, this effect can be understood as follows. In the topological regime, as the gap closes, the carriers' band-edge effective mass gets smaller, and $v_f$ gets larger. The opposite occurs when the energy gap opens (Fig.1(a,b)). When only point-like defects are considered in the material, $v_f$ determines the behavior of the conductivity. Therefore, a magnetic-field-induced gap closure causes a decrease in the resistance, whereas a gap opening causes an increase in resistance.

In order to plot the MR versus B using Eq.(7), a knowledge of $\Delta, v_z$ and $\widetilde{m}$ is required. The valley degeneracy and anisotropy of IV-VI materials also need to be accounted for. When B||[1-10], the Fermi surface consists of two ellipsoidal valleys having their major axis tilted by θ=90°, and two others tilted by θ=35° with respect to B (Fig.S2). [21] [20] $\Delta$ and $v_z(\theta)$ can be obtained from previous magnetooptical measurements. [24] Based on previous measurements of $\widetilde{m}$, we can determine $\widetilde{m}(\theta)$. [21] [24] The parameters for the angles of interest are shown in Table I.

| Pb$_{1-x}$Sn$_x$Se | n$_{SdH}$[pervalley] | $|\Delta|$[meV] | $v_z$ [10$^5$ m/s] (35°, 90°) | $\widetilde{m}/m_0$ (35°, 90°) | B$_c$ [T] (35°, 90°) |
|---|---|---|---|---|---|
| x=0.14 | 2x10$^{16}$ cm$^{-3}$ | 10 | 5.0, 4.8 | 0.20±0.02, 0.25±0.03 | N/A |
| x=0.19 | 6x10$^{16}$ cm$^{-3}$ | 10 | 4.8, 4.6 | 0.20±0.02, 0.25±0.03 | 17±2, 22±3 |

Table I. Parameters used to compute the MR shown in Fig. 3(d). The carrier density is determined from SdH measurements shown in Fig. 2. $B_c = \Delta \widetilde{m}/\hbar e$ is the field at which the N=0 LLs cross.

We now compute the variation of the N=0 conduction and valence LL as a function of magnetic field for both valleys for x=0.19 (Fig.3(c)), and calculate the MR using Eq.(7) for x=0.14 and x=0.19. In the trivial case for x=0.14, the MR is positive (Fig.3(d)), in agreement with the predictions of Goswami et al. for point defects [40] and with our data (Fig.2(a)). In the topological regime for x=0.19, the model yields a negative MR when $\Delta \widetilde{m}/2\hbar e < B < \Delta \widetilde{m}/\hbar e$ for each valley, (Fig.3(d)). We get a NLMR between 11T and 22T for the 90° valley and between 8.5T and 17T for the 35° valley. Two MR minima are thus expected at 22T and 17T. Experimentally, we observe a wide MR minimum at around B$_c$=20T (Fig.2(a)). The model thus agrees quantitatively with both the sign of the MR and position of the MR minimum. The broadening of the minumum can be due to the coexistence of the two minima resulting from valley degeneracy [20] and an anticrossing of N=0 LL (dashed line in Fig.3(c)) near B$_c$. [44]

The experimental onset of the NLMR is 5T. The model predicts an onset of about 8.5T. The onset calculated in the model is, however, non-universal and strongly depends on carrier population of different valleys. [45] For simplicity, a constant carrier population of valleys is assumed, leading to Eq. (4). This is not always the case in IV-VI TCIs thin films grown on BaF$_2$ since the N=0 LL disperse differently for different valleys and since a slight energy offset between different valleys may occur at low temperatures due to the mismatch of the expansion coefficients of the epilayers and the substrate. This causes a depopulation of one type of valleys and a repopulation of the other. [45] The most populated valley then dictates the behavior in the quantum limit, however, the carrier density in this valley will no longer be constant, resulting in a violation of Eq. (4). The onset of the NLMR will no longer be governed by the condition $\Delta \widetilde{m}/2\hbar e = B$ as inferred from Eq. (7) and will only be governed by the system entering the quantum limit.

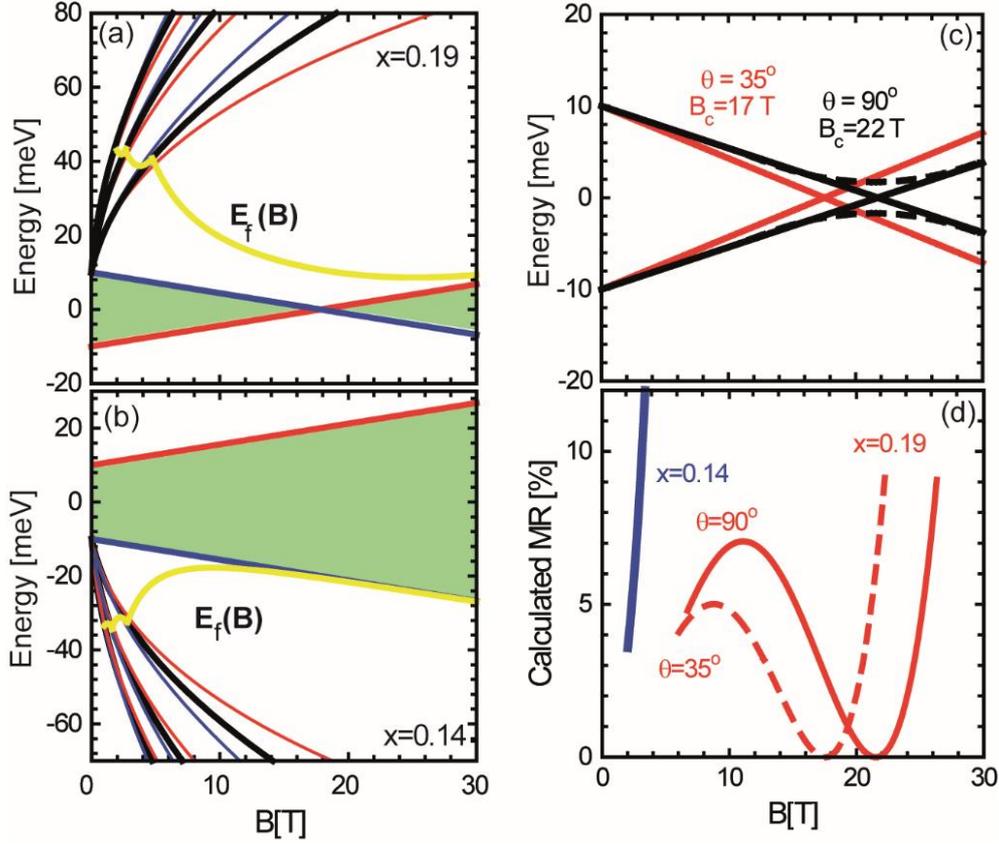

FIG 3. (Color Online). Massive Dirac LL (black) and spin-split LLs (red and blue) of $Pb_{1-x}Sn_xSe$ versus magnetic field for x=0.19 (a) and x=0.14 (b). The energy gap is $2\Delta=20meV$ and $v_z=4.8\times10^5$m/s for x=0.19 and $5.0\times10^5$m/s for x=0.14. $\tilde{m}=0.20 m_0$ is used for both samples. [24] $E_f$ versus magnetic field is shown in yellow. The energy gap is shaded in green. (c) N=0 LLs for the 35° and 90° valleys computed using the parameters in Table I. (d) MR calculated using Eq.(6) for parameters shown in table I, for x=0.14 and x=0.19 above the quantum limit.

Finally, the magnitude of the simulated MR is smaller than what is observed experimentally due to the rescaling of the MR by R(B=0), assumed to be given by Eq.(5) at B=0. Nevertheless, the shape of the NLMR, and its minimum agree very well with our model, without the use of any fit parameters. Most importantly, the model elucidates that the NLMR is observed in topologically non-trivial samples, and absent in trivial ones, as solely determined by the behavior of the N=0 LLs. A similar effect may occur in Dirac and Weyl semimetals when Zeeman splitting shifts the N=0 level in energy at high magnetic field. [46] [40] In this situation, a NLMR may be observed even if the Fermi energy is located far away from the Weyl nodes, and the chirality is not well-defined. [47]

In conclusion, we have shown that NLMR results from the anomalous behavior of the lowest bulk LL of topological materials (Fig.1). This MR and its anisotropy [20] are not qualitatively different from what is observed in Dirac and Weyl semimetals, as it only appears for B parallel to I. [48] [49] [50] [51] However, its origin is fundamentally different and is not related to the chiral anomaly. It is a result of the topologically non-trivial nature of bulk bands, the anomalous behavior of the N=0 Landau level and is a direct consequence of the inverted band structure of topological materials. Our results establish that NLMR is a hallmark of the topological insulating state, and can reconcile controversial interpretations of axial anomalous-like [52] [53] NLMR in candidate topological insulators such as, $ZrTe_5$, [23] [34] [54] and possibly $Pb_{0.75}Sn_{0.25}Te$ under pressure. [55] It may even be extended to the quasi-classical regime to explain the occurence of NLMR in $Bi_2Se_3$. [56]

**Acknowledgements:** We acknowledge A. Ernst, O. Pankratov, S. Tchoumakov, and M. Goerbig for useful comments. We also thank G. Bastard for several fruitful discussions. This work is supported by Agence

Nationale de la Recherche LabEx grant ENS-ICFP (ANR-10-LABX-0010/ANR-10-IDEX-0001-02 PSL) and by the Austrian Science Fund, Project SFB F2504-N17 IRON. TP acknowledges support from the Mahidol Wittayanusorn Scholarship and the Franco-Thai Scholarship. We also acknowledge the support of the HLD-HZDR, member of the European Magnetic Field Laboratory (EMFL).

Supplementary materials for:
# Negative longitudinal magnetoresistance from anomalous N=0 Landau level in topological materials

B.A. Assaf[1], T. Phuphachong[2], E. Kampert[3], V.V. Volobuev[4,5], P.S. Mandal[6], J. Sánchez-Barriga[6], O. Rader[6], G. Bauer[4], G. Springholz[4], L.A. de Vaulchier[2], Y. Guldner[2]

[1] Département de Physique, Ecole Normale Supérieure, PSL Research University, 24 rue Lhomond, 75005 Paris, France

[2] Laboratoire Pierre Aigrain, Ecole Normale Supérieure, PSL Research University, Université Pierre et Marie Curie, Sorbonne Universités, Université Denis Diderot, Sorbonne Cité, 24 rue Lhomond, 75005 Paris, France

[3] Dresden High Magnetic Field Laboratory (HLD-EMFL), Helmholtz-Zentrum Dresden-Rossendorf, 01328 Dresden, Germany

[4] Institut für Halbleiter und Festkörperphysik, Johannes Kepler Universität, Altenberger Straße 69, 4040 Linz, Austria

[5] National Technical University "Kharkiv Polytechnic Institute", Frunze Str. 21, 61002 Kharkiv, Ukraine

[6] Helmholtz-Zentrum Berlin für Materialien und Energie, Albert-Einstein Str. 15, 12489 Berlin, Germany

## S1. N=0 Landau level from Mitchell and Wallis Hamiltonian

Using the Mitchell and Wallis (1966) 6-band **k.p** formalism, we can write the following matrix Hamiltonian [1] [2]:

$$
\begin{array}{c c}
& \begin{array}{cccc} c^{1/2} F_n & v^{-1/2} F_{n+1} & c^{-1/2} F_{n+1} & v^{1/2} F_n \end{array} \\
\begin{array}{c} c^{1/2} F_n \\ v^{-1/2} F_{n+1} \\ c^{-1/2} F_{n+1} \\ v^{1/2} F_n \end{array} &
\begin{bmatrix}
\Delta + \hbar\tilde{\omega}\left(n+\frac{1}{2}\right) + \frac{1}{2}\tilde{g}\mu_B B & \sqrt{2e\hbar v_c^2 (n+1)B} & 0 & \hbar v_z k_z \\
\sqrt{2e\hbar v_c^2 (n+1)B} & -\Delta - \hbar\tilde{\omega}\left(n+\frac{3}{2}\right) + \frac{1}{2}\tilde{g}\mu_B B & \hbar v_z k_z & 0 \\
0 & \hbar v_z k_z & \Delta + \hbar\tilde{\omega}\left(n+\frac{3}{2}\right) - \frac{1}{2}\tilde{g}\mu_B B & \sqrt{2e\hbar v_c^2 (n+1)B} \\
\hbar v_z k_z & 0 & \sqrt{2e\hbar v_c^2 (n+1)B} & -\Delta - \hbar\tilde{\omega}\left(n+\frac{1}{2}\right) - \frac{1}{2}\tilde{g}\mu_B B
\end{bmatrix}
\end{array}
$$

In this formalism, only the conduction and valence band edge interaction are accounted for exactly, and 4 other far-bands intervene perturbatively. The following simplifications are used here to simplify far-band effective mass and g-factor contributions:

$$\overline{\omega}_c^l = -\overline{\omega}_v^l = \tilde{\omega} = \frac{eB}{\tilde{m}}$$

$$\overline{g}_c^l = -\overline{g}_v^l = \tilde{g}$$

The terms are defined in ref. [1]. $v_c$ is the critical Fermi velocity corresponding to each respective valley and $v_z = \dfrac{P_z}{m_0}$ is the Fermi velocity is the z-direction (the direction of the applied field), given by the k.p matrix element in the z-direction ($P_\parallel$ in the case of the longitudinal valley).

The lowest Landau level N=0 (σ=-1/2) is obtained by solving the inner block Hamiltonian for *n = - 1*.

$$\begin{array}{c c c c c}
 & c^{1/2}F_{-1} & v^{-1/2}F_{0} & c^{-1/2}F_{0} & v^{1/2}F_{-1}
\end{array}$$

$$\begin{array}{c}
c^{1/2}F_{-1} \\
v^{-1/2}F_{0} \\
c^{-1/2}F_{0} \\
v^{1/2}F_{-1}
\end{array}
\begin{bmatrix}
\Delta - \hbar\tilde{\omega}\left(\frac{1}{2}\right) + \frac{1}{2}\tilde{g}\mu_B B & 0 & 0 & \hbar v_z k_z \\
0 & -\Delta - \hbar\tilde{\omega}\left(\frac{1}{2}\right) + \frac{1}{2}\tilde{g}\mu_B B & \hbar v_z k_z & 0 \\
0 & \hbar v_z k_z & \Delta + \hbar\tilde{\omega}\left(\frac{1}{2}\right) - \frac{1}{2}\tilde{g}\mu_B B & 0 \\
\hbar v_z k_z & 0 & 0 & -\Delta + \hbar\tilde{\omega}\left(\frac{1}{2}\right) - \frac{1}{2}\tilde{g}\mu_B B
\end{bmatrix}$$

The inner block reduces to:

$$\begin{bmatrix}
-\Delta - \frac{\hbar\tilde{\omega}}{2} + \frac{1}{2}\tilde{g}\mu_B B & \hbar v_z k_z \\
\hbar v_z k_z & \Delta + \frac{\hbar\tilde{\omega}}{2} - \frac{1}{2}\tilde{g}\mu_B B
\end{bmatrix}$$

We can now solve the eigenvalue problem for the N=0 (σ= - 1/2) level:

$$\begin{vmatrix}
-\Delta - \frac{\hbar\tilde{\omega}}{2} + \frac{\tilde{g}\mu_B B}{2} - E & \hbar v_z k_z \\
\hbar v_z k_z & \Delta + \frac{\hbar\tilde{\omega}}{2} - \frac{\tilde{g}\mu_B B}{2} - E
\end{vmatrix} = 0$$

We get the following equation:

$$\left(E - \frac{\hbar\tilde{\omega}}{2} + \frac{\tilde{g}\mu_B B}{2} - \Delta\right)\left(E + \frac{\hbar\tilde{\omega}}{2} - \frac{\tilde{g}\mu_B B}{2} + \Delta\right) - (\hbar v_z k_z)^2 = 0$$

The energy eigenvalue to the lowest Landau level now rewritten as $E_0$ is then given by:

$$E_0 = \pm\sqrt{\left(\Delta + \frac{\hbar\tilde{\omega}}{2} - \frac{\tilde{g}\mu_B B}{2}\right)^2 + (\hbar v_z k_z)^2}$$

For the following we will only consider the N=0 Landau level of the conduction band with the (+) sign, keeping in mind that the corresponding one of the valence band is simply given by its opposite. Thus,

$$E_0 = \sqrt{\left(\Delta + \frac{\hbar\tilde{\omega}}{2} - \frac{\tilde{g}\mu_B B}{2}\right)^2 + (\hbar v_z k_z)^2}$$

When going through the topological phase transition, Δ changes sign yielding:

$$E_0 = \sqrt{\left(\pm\Delta + \frac{\hbar\tilde{\omega}}{2} - \frac{\tilde{g}\mu_B B}{2}\right)^2 + (\hbar v_z k_z)^2}$$

Or,

$$E_0 = \sqrt{\left(|\Delta| \pm \left[\frac{\hbar\tilde{\omega}}{2} - \frac{\tilde{g}\mu_B B}{2}\right]\right)^2 + (\hbar v_z k_z)^2} \qquad \text{(+ for trivial, – for topological)}$$

We make an even further simplifying assumption:

$$\tilde{g}\mu_B B \approx -\hbar\tilde{\omega}.$$

This is justified since it can be shown that $|\tilde{g}\mu_B B| \approx |\hbar\tilde{\omega}|$. [3][4] For PbSe, both parameters have been reported by Pascher et al. [4] for the conduction band of the longitudinal valley: $\bar{g}_c^l == \tilde{g} = -7.5 \pm 1$ and $m_c^t = \tilde{m} = (0.27 \pm 0.05)m_0$. This gives:

$$\bar{g}_c^l \mu_B B \approx (0.435 \pm 0.06 meV/T)B \text{ and } \frac{\hbar eB}{m_c^t} \approx (0.43 \pm 0.07 meV/T)B$$

The findings of Pascher et al. confirm a Zeeman splitting due to far-bands almost equal to far-band mass contributions to the cyclotron energy. According to Calawa et al. [7], far-band contributions vary by less than 10% as a function of Sn content up to x= 0.28, and thus we use this fact for our analysis of Pb$_{1-x}$SnSe, too. The energy eigenvalue of the lowest Landau level is then given by:

$$E_0 = \pm\sqrt{(\Delta + \hbar\tilde{\omega})^2 + (\hbar v_z k_z)^2}$$

### S2. Zeeman splitting and the effective g-factor in IV-VI semiconductors

Typically, the Zeeman splitting energy is the energy difference between N$^+$ and N$^-$ levels. As illustrated in Fig. S1(a), in the massive Dirac model that includes spin-splitting, this would simply imply the lifting of the degeneracy of the N levels due the far-band g-factor. Bear in mind, however, that for an ideal massive Dirac model (2-band **k.p**, Fig. S1(b)), the far-band contribution to the g-factor is equal to zero and the Landau levels are spin degenerate. Thus the spin-splitting is equal to the cyclotron energy, yielding for x=0.19:

$$g = \frac{2m_0}{m} \approx \frac{2}{0.0084} \approx 238$$

(For free electrons $m = m_0$ and $g = 2$)

Here we have used $\frac{m_0}{m} = \frac{v_c^2 m_0}{\Delta} = \frac{P_\perp^2}{\Delta m_0}$ to compute the g-factor as given by Pascher et al. [4] without far-band terms. Such large g-factors, and the non-spin-degenerate character of the Landau levels, including the lowest one, are common in narrow gap systems such as Pb$_{1-x}$Sn$_x$Se,[4] Pb$_{1-x}$Sn$_x$Te and Hg$_{1-x}$Cd$_x$Te. [5]

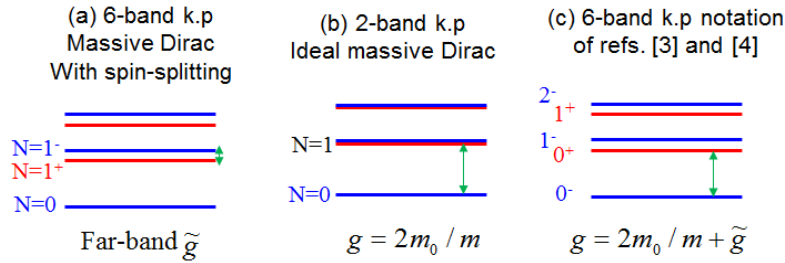

Figure S1. (a) Massive Dirac Landau levels with spin-splitting from 6 band **k.p** model. (b) Ideal massive Dirac model from 2 band k.p treatment. (c) 6 band **k.p** model with levels indexed according Refs . [4]

Therefore, in a massive Dirac model that includes spin splitting, both $g$ and the far-band g-factor $\tilde{g}$ need to be included. This subtlety can be better understood when the notation of Melngailis et al.[3] and Pascher et al.[4] is used to index the Landau levels. It can be easily seen in Fig. S1(c) that the actual g-factor for the longitudinal valley, contains both contributions and is written as [4]:

$$g = \frac{2m_0}{m} + \tilde{g}$$

For x=0.19:
$$g = \frac{2}{0.0084} - 7.5 = 230.5$$

## S3. Far-band mass correction anisotropy

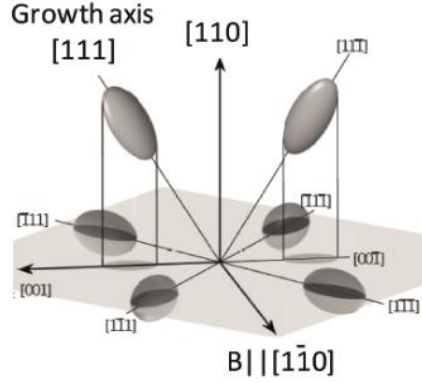

Figure S2. Fermi surface ellipsoids of $Pb_{1-x}Sn_xSe$ at the L-points of the bulk Brillouin zone. The magnetic field applied along [1-10] direction is shown.

In our previous work on magnetooptical characterization of the band structure of $Pb_{1-x}Sn_xSe$, we measured $\tilde{m} \approx (0.24 \pm 0.03)m_0$ for the oblique valleys in the Faraday geometry ($\theta=71°$). [6] From Calawa et al. we have $\tilde{m} \approx (0.22)m_0$ for the [001] valleys having ($\theta=53°$). [7] Using these two experimental results and the fact that:

$$\tilde{m}(\theta) \approx m_t \left[ \frac{\sin^2(\theta) m_t}{m_l} + \cos^2(\theta) \right]^{-1/2}$$

we find the anisotropy ratio for far-band contributions:

$$\frac{m_\ell}{m_t} \approx 1.7 \pm 0.1,$$

and estimate $m_t \approx (0.19 \pm 0.02)m_0$ and $m_l \approx (0.32 \pm 0.05)m_0$. Here $m_t$ and $m_l$ are the transverse and longitudinal far-band mass correction terms as defined in ref. 1. We can now compute $\tilde{m}(\theta)$ for any angle. When B||[1-10], it can easily be shown that two oblique valleys (the [1-11] and the [-111] valleys) are tilted by 35° with respect to the magnetic field and one longitudinal ([111]) and one oblique valley ([11-1]) are tilted by 90°.

$$\tilde{m}(90°) \approx m_t \left[ \frac{\sin^2(\theta) m_t}{m_l} + \cos^2(\theta) \right]^{-1/2} \approx (0.25 \pm 0.03)m_0$$

$$\tilde{m}(35°) \approx m_t \left[ \frac{\sin^2(\theta) m_t}{m_l} + \cos^2(\theta) \right]^{-1/2} \approx (0.2 \pm 0.02)m_0$$

## S4. Hall effect

| For DC fields | $n_1$ [cm$^{-3}$] | $n_2$ [cm$^{-3}$] | $\mu_1$ [cm$^2$/Vs] | $\mu_2$ [cm$^2$/Vs] | $\rho_{tot}$ [$\Omega$.m] |
|---|---|---|---|---|---|
| x=0.10 | **3.0 x 10$^{17}$ elec.** | 7 x 10$^{17}$ holes | **50000** | 1500 | 4x10$^{-7}$ |
| x=0.14 | **1.5 x 10$^{17}$ holes** | 6 x 10$^{17}$ holes | **60000** | 1600 | 4x10$^{-6}$ |
| x=0.19 | **3.5 x 10$^{17}$ elec.** | 1 x 10$^{18}$ holes | **17000** | 500 | 1x10$^{-6}$ |
| x=0.23 | **3 x 10$^{16}$ elec.** | 8 x 10$^{17}$ holes | **40000** | 2000 | 2 x 10$^{-5}$ |
| x=0.30 | **4.5 x 10$^{17}$ holes** | 9 x 10$^{17}$ holes | **8500** | 4500 | 3 x 10$^{-5}$ |

Table S1. Transport parameters determined from the Hall effect using a 2 parameter Drude fit. Note that all samples show two carrier transport. The low mobility channel is either due to a 2D Fermi surface channel or an interfacial layer. The bulk is found to have a low carrier density n$_1$ and a high mobility $\mu_1$ in all samples as confirmed by the low field Shubnikov-de-Haas oscillations and by magnetooptical data. [6]

## S5. Quantum oscillations in x=0.23 and x=0.3

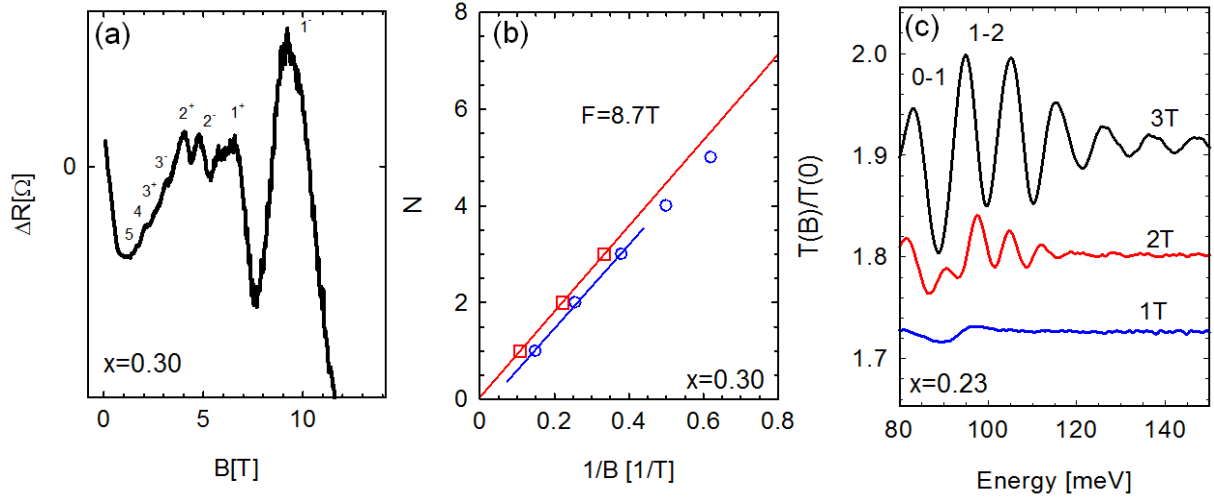

Figure S3. (a) Shubnikov-de-Haas oscillations in x=0.3 at 10K. (b) Landau index versus 1/B extracted from (a). Red and blue indicate the series of opposite spin as indexed in (a). (c) Magnetooptical infrared absorption spectra taken between 1T and 3T for x=0.23 at 4.5K. The N=0-N=1 interband transition is already visible and strong in amplitude at 2T.

Shubnikov-de-Haas data shown is Fig. S3(a) for x=0.3. Spin is taken into account in the plot of N versus 1/B as shown Fig. S3(b) similarly to ref. [3]. The lowest N=1 Landau level crosses the Fermi energy close to 9T. For x=0.23, magnetooptical Landau level spectroscopy measurements at low fields are performed. The details of these measurements are shown in ref. [4] The N=0 to N=1 interband Landau level transition is observed at 2T and above indicating that the N=1 Landau level crosses the Fermi energy at 2T (Fig.S3(c)).

The Landau level energies for Pb$_{1-x}$Sn$_x$Se derived from the Mitchell and Wallis Hamiltonian for N>0 has been discussed in our previous work [6] [8] and is given by:

$$E_{N>0}^{c,\pm} = \mp\hbar\widetilde{\omega} + \sqrt{\Delta^2 + 2v_D^2\hbar eBN}$$

$$E_{N>0}^{v,\pm} = -E_{N>0}^{c,\pm}$$

$\Delta$ denotes the energy gap divided by two (E$_g$/2), $v_D$ is the Dirac velocity, and $\widetilde{\omega} = eB/\widetilde{m}$ as discussed in the main text. When spin splitting is neglected ($\widetilde{\omega} = 0$), we recover a classical massive Dirac dispersion. The Landau level dispersion versus B is plotted for x=0.23 and x=0.30 in Fig. S4.

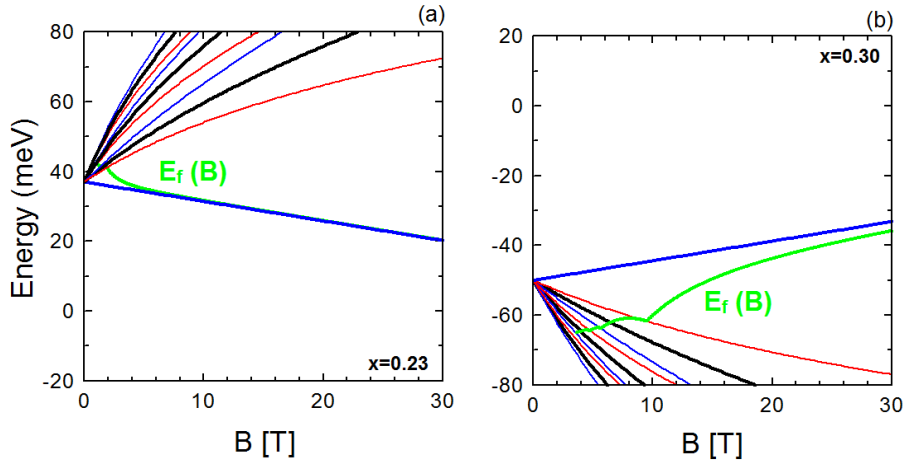

FIG S4. Landau level dispersion versus magnetic field for x=0.23 (a) and x=0.3 (d), with the field dependence of the Fermi energy. Landau levels from an ideal massive Dirac model (black) and a massive Dirac model that includes spin are also shown (red, blue). For N>0, spin splitting is relevant and taken into account for x=0.3, it is rather small in the other samples.

## S6. MR anisotropy

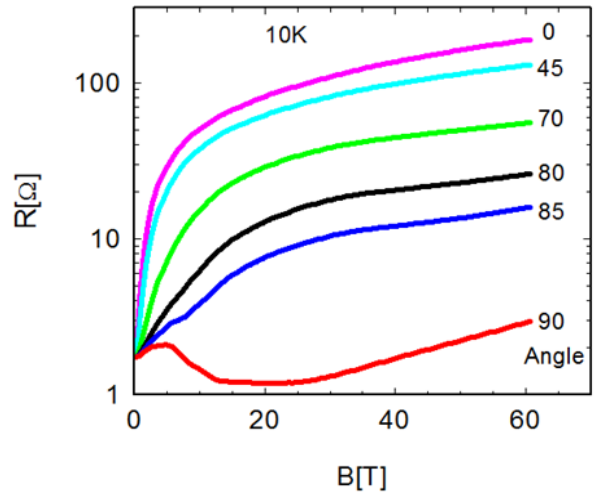

Figure S5. (a) Resistance as a function of magnetic field for different angles θ, up to 60T at 10K for x=0.19. θ is the angle between the applied field and the normal. θ=0° corresponds to an out-of-plane magnetic field, normal to the sample surface (B//[111]). When θ=90° the magnetic field is aligned parallel to the current. The MR is highly anisotropic and positive MR is restored already for θ=85°.

## S7. Determination of the bulk band gaps of $Pb_{1-x}Sn_xSe$

ARPES measurements were performed to identify that our samples with a Sn content corresponding to x>0.16 exhibit a gapless Dirac cone with a linear dispersion at the $\overline{\Gamma}$ -point of the surface Brillouin zone as a sign for their topological character. The precise determination of the bulk band gaps from the ARPES data in the present $Pb_{1-x}Sn_xSe$ samples is a rather challenging task, since by varying the photon energy (i.e., momentum perpendicular to the surface) we do not observe distinct intensity contributions from the bulk conduction band (BCB) in the inverted regime. This is in contrast to the contribution from the surface state which gives a much higher photoemission intensity. Note that the exact determination of the bulk band gaps using ARPES requires a variation of the wave vector perpendicular to the surface with extreme precision, so that one perfectly "slices" the edge of the BCB (as far as it is occupied) and the one of the bulk valence band (BVB) with an overall accuracy substantially better than the size of the energy gap in these narrow gap systems

We instead determine the bulk band gaps of our $Pb_{1-x}Sn_xSe$ samples from the investigation of magnetooptical interband transitions in Faraday geometry (B// [111]). This is discussed in detail in ref. [6] for samples x=0.14 and x=0.19, the same samples that were measured and analyzed in detail in this work.